\documentclass[sigconf,nonacm]{acmart}

\usepackage{graphicx}
\usepackage{hyperref}
\usepackage{siunitx}
\usepackage{caption}
\usepackage{subcaption}
\usepackage{float}
\usepackage{xcolor}
\AtBeginDocument{%
  \providecommand\BibTeX{{%
    \normalfont B\kern-0.5em{\scshape i\kern-0.25em b}\kern-0.8em\TeX}}}

\renewcommand\footnotetextcopyrightpermission[1]{} 

\begin{document}

\title{A Fair and Comprehensive Comparison of Multimodal Tweet Sentiment Analysis Methods}





\author{Gullal S. Cheema$^1$, Sherzod Hakimov$^1$, Eric M{\"u}ller-Budack$^1$ and Ralph Ewerth$^{1,2}$}
\affiliation{%
  \institution{$^1$TIB -- Leibniz Information Centre for Science and Technology \\ $^2$L3S Research Center, Leibniz University Hannover}
  \city{Hannover}
  \country{Germany}}
\email{{gullal.cheema, sherzod.hakimov, eric.mueller, ralph.ewerth}@tib.eu}

\renewcommand{\shortauthors}{Cheema et al.}

\begin{abstract}
Opinion and sentiment analysis is a vital task to characterize subjective information in social media posts.
In this paper, we present a comprehensive experimental evaluation and comparison with six state-of-the-art methods, from which we have re-implemented one of them. 
In addition, we investigate different textual and visual feature embeddings that cover different aspects of the content, as well as the recently introduced multimodal CLIP embeddings. 
Experimental results are presented for two different publicly available benchmark datasets of tweets and corresponding images. 
In contrast to the evaluation methodology of previous work, we introduce a reproducible and fair evaluation scheme to make results comparable. Finally, we conduct an error analysis to outline the limitations of the methods and possibilities for the future work.
\end{abstract}

\renewcommand{\authors}{Gullal S. Cheema, Sherzod Hakimov, Eric M{\"u}ller-Budack, Ralph Ewerth}
\keywords{Multimodal Sentiment Analysis, Information Retrieval, Social Media, Computer Vision, Natural Language Processing, Transformer Models}

\maketitle
\pagestyle{empty}

\section{Introduction}
Social media has become a phenomenon in terms of its usage by the general public, traditional media, enterprises, and also as a forum for discussing research in academia. With the evolution of the Internet, social media sites, in particular, have become multimodal in nature with content including text, audio, images, and videos to engage different senses of a user. Similarly, sentiment analysis techniques have also progressed from extensively explored text-based~\cite{liu2012survey,medhat2014sentiment} to multimodal sentiment analysis~\cite{soleymani2017survey} of image-text pairs or videos. With two or more modalities, the problem becomes more challenging since every modality might differently influence the overall sentiment, and modalities can have a complex interplay. For image-text pairs, this is even harder as images are perceived as a whole, whereas text is read sequentially. Existing approaches focus on different types of features~\cite{niu2016sentiment,xu2017multisentinet} and complex attention mechanisms~\cite{jiang2020fusion,xu2018co} to capture the inter-dependencies between image and text to build multimodal models.


Psychological studies have found that human visual attention generally prioritizes emotional content over non-emotional content~\cite{brosch2010perception,compton2003interface}. A recent study by Fan \textit{et al.}~\cite{Fan_2018_CVPR} evaluated the inter-relationships of image sentiment and visual saliency in deep convolutional neural network (CNN) models. They proposed a model that prioritizes emotional objects over other objects to predict sentiment, just like human perception. It indicates that to learn a multimodal model for sentiment prediction, visual features should contribute and consider different objects, facial expressions, and other salient regions in the image. Besides, to learn a multimodal model for sentiment detection, extracted features from two modalities need to be combined in a way that reflects the overall sentiment of the image-text pair. Even though the existing approaches~\cite{xu2017multisentinet,xu2018co,jiang2020fusion} for image-text sentiment detection proposed complex attention mechanisms over different types of features, they fall short on the analysis of features, the number of visual features used, and lack a reproducible evaluation, which hampers the progress in this field.

In this paper, we study the impact of different visual features in combination with contextual text representations for multimodal tweet sentiment classification and present a comprehensive comparison with six state-of-the-art methods. In contrast to previous work, we investigate four different visual feature types: facial expression, object, scene, and affective image content. We utilize a simple and efficient multimodal neural network model (Se-MLNN: Sentiment Multi-Layer Neural Network) that combines several visual features with contextual text features to predict the overall sentiment accurately. 
In our experiments, we also test the recently proposed \emph{CLIP} model (contrastive Language-Image Pre-training~\cite{radford2021learning}), which is specifically trained on millions of image-text pairs and reports impressive zero-shot performance on image classification datasets like \emph{ImageNet}~\cite{russakovsky2015imagenet} and \emph{Places365}~\cite{zhou2017places}. We use this model instead of pre-trained multimodal transformers due to the volume and variety of data it exploited for pre-training, which makes it attractive for different kinds of visual recognition tasks.

We use the publicly available benchmark MVSA~\cite{niu2016sentiment} (Multi-View Social Data) that consists of two different datasets of tweets and corresponding images. We provide a detailed analysis of both image and text features complemented with an extensive experimental study and outline the limitations of existing approaches. All of the existing approaches for the MVSA datasets use randomly generated, unpublished train and test splits, which makes it impossible to reproduce results or fairly compare them. Thus, we apply k-fold cross-validation so that every sample in datasets is tested once. We share the source code and the new dataset splits used in this paper\footnote{\url{https://github.com/cleopatra-itn/fair_multimodal_sentiment}}.




\section{Related Work}\label{sec:related_work}
Sentiment detection has been extensively explored for textual social media data, with earlier approaches that were lexicon-based evolving to statistical and machine learning-based classification in the last decade. \emph{SentiStrength}~\cite{thelwall2010sentiment} is a well-known lexicon-based approach for short text built using widely occurring words and phrases on social media. Later, Saif \textit{et al.}~\cite{saif2016contextual} developed \emph{SentiCircles} specifically for Twitter sentiment analysis by taking into account the co-occurrence of words in different contexts in tweets. With the prevalence of deep learning, convolutional neural networks (CNNs)~\cite{alayba2018combined,kim2014convolutional} and sequential models like Long Short-term Memory (LSTM)~\cite{huang2016modeling} networks have been successfully used for tweet sentiment classification. Shin \textit{et al.}~\cite{shin2016lexicon} developed a hybrid approach by integrating lexicons with a CNN using an attention mechanism. With the rise of image and video data on social media sites like Instagram, Flickr, and Twitter, visual sentiment analysis has attracted a lot of attention recently. Also, in this case, the techniques can be broadly divided into mid-level and deep learning representations. Borth \textit{et al.}~\cite{borth2013sentibank,borth2013large} developed \emph{SentiBank}, a library to detect visual concepts corresponding to \num{1200} adjective noun pairs (ANP) for sentiment classification. The concept classifiers rely on low-level visual representations like the ``gist'' descriptor~\cite{oliva2001modeling} and wavelet-like Haar features~\cite{viola2001rapid}. Yuan \textit{et al.}~\cite{yuan2013sentribute} detected the presence of faces and used facial expressions in combination with low-level visual features to predict the sentiment in images. Using the expressiveness of pre-trained CNNs, You \textit{et al.}~\cite{you2015robust} fine-tuned the networks for binary sentiment classification on Flickr and Twitter image datasets. Later, they extended the visual sentiment problem to affective image content analysis~\cite{you2016building} to predict emotions like \textit{amusement, anger, awe, fear, sadness} and \textit{excitement}. Recently, this problem has been explored by several researchers~\cite{yang2018weakly,yang2017joint,zhan2019zero,zhao2019pdanet,zhu2017dependency} and is discussed in a comprehensive survey by Zhao \textit{et al.}~\cite{zhao2018affective}.

Multimodal sentiment analysis for social media has gained popularity due to the challenge of combining information from two or more modalities that influence sentiment. Cao \textit{et al.}~\cite{cao2016cross} extracted low-level visual representations, \emph{SentiBank} visual concepts, and lexicon-based features from text to predict sentiment using late fusion strategies. To capture high-level concepts in both image and text, Cai \textit{et al.}~\cite{cai2015convolutional} as well as Yu \textit{et al.}~\cite{yu2016visual} trained a shallow CNN for text and a deep CNN for images with shared representation to predict sentiment and achieved much better performance than the previous approaches. For the MVSA~\cite{niu2016sentiment} dataset, in particular, several techniques improved the performance by focusing on cross-modal representations that capture the influence of modalities towards the sentiment. Xu \textit{et al.}~\cite{xu2017multisentinet} proposed an approach using deep CNN with pre-trained features that encode object and scene information from images and aggregated it with contextual \emph{GloVe}~\cite{pennington2014glove} (Global Vectors for Word Representation) word embeddings from the text. They used visual feature-guided attention to capture the influence of visual features over word embeddings instead of simply concatenating them for predicting the sentiment. Later, Xu \textit{et al.}~\cite{xu2018co} proposed a co-memory attention mechanism using similar features to capture the interaction between two modalities and their influence on the sentiment. Recently, Jiang \textit{et al.}~\cite{jiang2020fusion} proposed another attention mechanism where they used both cross-modal attention fusion followed by modality-specific CNN-gated feature extraction to learn a better representation. They used \emph{ImageNet}~\cite{russakovsky2015imagenet} pre-trained \emph{ResNet}~\cite{he2016deep} for visual features, and experimented with GloVe~\cite{pennington2014glove} and \emph{BERT}~\cite{devlin2018bert} (Bidirectional Encoder Representations from Transformers) embeddings for textual features to achieve state-of-the-art results for the MVSA dataset.

\section{Multimodal Sentiment Classification}
\label{sec:approach}

The main idea is to exploit and investigate different kinds of high-level visual features and combine them with a textual model. The use of channel features as a sequence in conjunction with word embeddings by previous approaches limits their model to two modalities or types of features~\cite{jiang2020fusion,xu2017multisentinet,xu2018co}. In contrast, we aim to investigate the impact of our suggested high-level visual features, which are objects, scenes or places, facial expressions, and the overall affective image content in a more efficient and less complex framework.

\begin{figure}[!ht]
	\centering
	\includegraphics[width=1.0\linewidth]{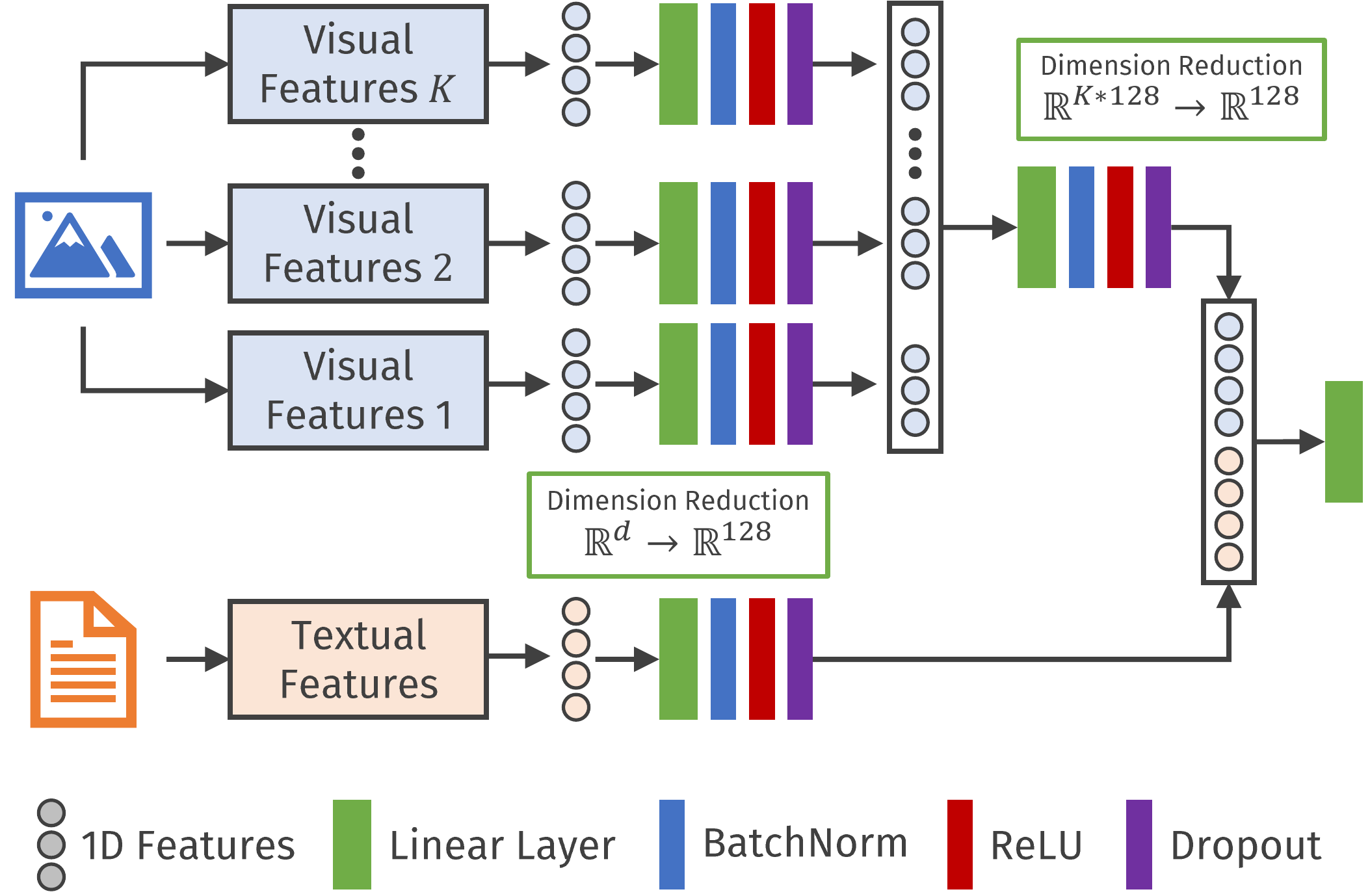}
	\caption{Se-MLNN: Proposed architecture for multimodal sentiment classification. Here $d$ in $\mathbb{R}^d$ is different for every feature and is provided in Section~\ref{sec:visual},~\ref{sec:textual} and~\ref{sec:multi}. Every feature irrespective of the dimension $d$ is projected down to 128 (First Layer) in order to keep the number of parameters low and not to introduce feature bias. The final layer ($\mathbb{R}^{256} \rightarrow \mathbb{R}^3$) is followed by softmax that outputs the probability of each sentiment.}
	\label{fig:model}
\end{figure}

A crucial difference between our approach and recent multimodal tweet sentiment approaches~\cite{jiang2020fusion,xu2018co} is that we use pooling to get one embedding per image instead of using channel features from a pre-trained CNN as a sequence. Similarly, we use a pooling strategy for getting one embedding per a tweet from a textual model. 
Another difference is that instead of relying on learning bi-attention weights from a limited amount of data, we use a multi-layer neural network to combine different features to influence the sentiment. To investigate the impact of our suggested high-level visual features, we propose a three-layer neural network, where the first two layers aggregate features from different modalities, and the third layer is used for the classification of sentiment. The architecture of our approach is shown in Figure~\ref{fig:model}. The training details are provided in Section~\ref{sec:experiments}. Next, we describe the individual models for each modality and their encoding process to understand the proposed approach. 

\subsection{Visual Features}\label{sec:visual}
\subsubsection{Object Features ($E_o$):}
Different objects in a picture can incite a particular sentiment in a person. For instance, a cute dog or flowers might bring a positive sentiment, whereas a snake may incite a negative sentiment depending on the context. 
To encode objects and the overall image content, we extract features from a pre-trained \emph{ResNet} model~\cite{he2016deep} trained on \textit{ImageNet}~\cite{russakovsky2015imagenet}. We use \emph{ResNet}-50 and its last convolution layer to extract features instead of the object categories (final layer). The final convolutional layer outputs 2048 feature maps each of size 7 x 7, which is then pooled with a global average to get a 2048-dimensional vector. 

\subsubsection{Place and Scene Features ($E_s$):}
A scene or a place can also incite different sentiments in a person. For instance, a candy store might bring a positive sentiment, whereas a catacomb might incite a negative sentiment depending on the context. 
To encode the scene information of an image, we extract features from a pre-trained \emph{ResNet}~\cite{he2016deep} model trained on \textit{Places365}~\cite{zhou2017places}. In this case, we use \emph{ResNet}-101 and follow the same encoding process as above.

\subsubsection{Facial Expressions ($E_f$):}
The presence of faces and facial expressions (smiling vs. sad face) in an image can also influence the sentiment in an observer. In the \emph{MVSA} dataset, we found that around $50\%$ of the images contain faces with an average of 2 to 3 faces per image. 
In order to encode information about facial expressions, we extract the final layer features from a pre-trained~\cite{wujie_face2018} \emph{VGG-19} (Visual Geometry Group) model~\cite{simonyan2014very} that is trained on around \num{28000}~\cite{face_exp2013,goodfellow2013challenges} face images based on the following seven classes: \textit{angry}, \textit{disgust}, \textit{fear}, \textit{happy}, \textit{sad}, \textit{surprise} and \textit{neutral}. 
Before extracting the expression features, we first detect faces from an image using a state-of-the-art DSFD~\cite{li2018dsfd} (Dual Shot Face Detector) face detector, which are then rescaled to $48 \times 48$ pixels and input to the \emph{VGG} network. 
For a given image, if the detector~\cite{li2018dsfd} detects $K$ faces, the \emph{VGG} network outputs $K$ 512-dimensional features that are averaged to get the final feature vector or vector with zeros if no faces are detected.

\subsubsection{Affective Image Content ($E_a$):} 
Overall affective image content can also be important for multimodal sentiment detection, and research in this area has made rapid progress in recent years with famous datasets from popular social media image sharing websites such as Flickr and Instagram. 
To encode the overall emotion, we first fine-tune a \emph{ResNet}-50 \textit{ImageNet} model on publicly available FI (Flickr \& Instagram) dataset~\cite{you2016building} and extract the last layer convolution features as described above for object and scene embeddings. The dataset consists of around \num{23000} training images and eight emotion classes: \textit{amusement}, \textit{anger}, \textit{awe}, \textit{contentment}, \textit{disgust}, \textit{excitement}, \textit{fear} and \textit{sadness}.

\subsection{Textual Features}\label{sec:textual}
Since context and meaning of the words are equally important for the influence of the whole sentence towards the sentiment, we use \emph{RoBERTa}-Base~\cite{liu2019roberta} (Robustly optimized BERT approach) to extract contextual word embeddings and employ different pooling strategies to get a single embedding for the tweet. 
We experimentally found that the average of the last four layers is the most useful and fixed that embedding for all our \emph{RoBERTa} embedding experiments. We finally take an average over the word embeddings to get the single text embedding of 768 dimensions for every tweet. For pre-processing text, we normalize text following~\cite{baziotis-pelekis-doulkeridis:2017:SemEval2} and conduct three experiments by keeping ($E_T^{+HT}$) and removing ($E_T^{-HT}$) the hashtags from the text and also on the raw tweet ($E_T^{RAW}$) text. 


\subsection{Multimodal Features}\label{sec:multi}
We use the recently proposed multimodal model \emph{CLIP}~\cite{radford2021learning} that is trained on 400 million image-text pairs collected from the Internet. The model is trained to predict which caption goes with which image and in doing so it learns expressive image representation without the need for millions of labeled training examples. In comparison to multimodal transformers~\cite{lu2019vilbert,lu2019vilbert,li2020oscar}, the model uses pairwise learning over n-pairs of image and text and does not use any cross-attention mechanism to learn multimodal features. This makes the model easy to use as image and text embeddings can be independently computed from the respective image and text encoders. Because of the variety and a large amount of data, the model shows competitive zero-shot recognition performance on 30 different computer vision datasets in comparison to their supervised baselines. This suggests that the amount and quality of visual information encoded in the visual features of the model are much better than the \emph{ImageNet} and \emph{Places365} supervised pre-trained models.

We use a publicly available \emph{CLIP} model\footnote{\url{https://github.com/openai/CLIP}} and a variant that has visual and textual transformer as image and text encoder backbones. We extract both image and text features from the model where both image ($C_I$) and text embeddings are 512-dimensional vectors. For text, we use the same pre-processing as used for textual models and this results in three types of text embeddings, with hashtags ($C_T^{+HT}$), without hashtags ($C_T^{-HT}$), and raw text ($C_T^{RAW}$).

\section{Experimental Setup and Results}\label{sec:experiments}

\subsection{Dataset and Training Details}
\subsubsection{Datasets} 
We use the \emph{MVSA-Single} (MVSA-S) and \emph{MVSA-Multiple} (MVSA-M) datasets \cite{niu2016sentiment} to test our model. The two datasets contain \num{4869} and \num{19598} image-text pairs from Twitter, where both image and text are annotated with a separate label by a single annotator (\emph{MVSA-S}) or three annotators (\emph{MVSA-M} with three annotations for each sample), respectively. 
For a fair comparison and to be consistent with previous work, we process to get the multimodal label and filter the two datasets according to~\cite{xu2017multisentinet,xu2018co}, which results in \num{4511} and \num{17025} image-text pairs, respectively. 
To summarize it, the majority of the assigned class labels over each pair is the pooled label, and the final label falls under three cases: 1) label is valid and same if both have the same label, 2) label is valid and a polar label if either is positive or negative and the other is neutral, and 3) tweet is a conflict and filtered if image and text have opposite polarity labels. The \emph{MVSA-Single} dataset consists of \num{470} \textit{neutral}, \num{2683} \textit{positive} and \num{1358} \textit{negative} samples, while the \emph{MVSA-Multiple} dataset consists of \num{4408} \textit{neutral}, \num{11318} \textit{positive} and \num{1299} \textit{negative} samples.

\subsubsection{Evaluation and Comparison} 
We conduct 10-fold cross-validation where every split ends up with 8:1:1 ratio data and the same label distribution in training, validation and test set. For our ablation study, we report the average accuracy and weighted F1 scores (according to class size) over the 10 folds in Table~\ref{tab:unimodal_results} and~\ref{tab:multimodal_results}. For comparison with two other approaches, we report minimum, maximum, and average accuracy and weighted-F1 measure for all our runs. Our results cannot be directly compared with some previous approaches' reported results (taken from Jiang \textit{et. al.}~\cite{jiang2020fusion}) and are only here for reference (marked with~$^{\dagger}$). We evaluate the publicly available \emph{MultiSentNet}~\cite{xu2017multisentinet} implementation\footnote{\url{https://github.com/xunan0812/MultiSentiNet}} and re-implemented the \emph{FENet}~\cite{jiang2020fusion} model to compare their results (marked with *) in Table~\ref{tab:comparison}.

\subsubsection{Training Details}
We use cross-entropy as an objective function and the Adam (adaptive moment estimation) ~\cite{kingma2014adam} optimizer for updating the neural network parameters. We observe that the \emph{MVSA-M} dataset label pooling strategy results in noisy labels and to mitigate that we use label smoothing so that the model does not become overconfident. With a smoothing factor of $\alpha = \num{0.1}$ and $K = 3$ classes, the one-hot encoded label vector $y$ becomes:
\begin{equation}
    y_{ls} = (1-\alpha) * y + \alpha/(K-1)
\end{equation}
The learning rate is set to \num{2e-5} and all the models are trained for \num{100} epochs. We decay the learning rate by a factor of 10 if the validation loss does not decrease for five epochs. A batch size of 32 and 128 is used for \emph{MVSA-S} and \emph{MVSA-M}, respectively. A dropout with the ratio of \num{0.5} is applied after all the intermediate linear layers to avoid over-fitting. We save the best model (of all epochs) according to the lowest validation loss while training and use it for testing. 
We use PyTorch\footnote{https://github.com/pytorch/pytorch} for our experiments and extract object features from its publicly available \emph{ImageNet} pre-trained \emph{ResNet-50}~\cite{he2016deep} model. We train a \emph{ResNet-101} model on the \emph{Places365} dataset and use it for extracting scene features.

\bgroup
\setlength{\tabcolsep}{6pt} 
\renewcommand{\arraystretch}{1.2} 
\begin{table}[!ht]
\caption{Unimodal and multiple visual feature results for \emph{MVSA-Single} and \emph{MVSA-Multiple}. Accuracy and F1 scores are averaged over 10 folds.}
\begin{tabular}{|l|c|c|c|c|}
\hline
\textbf{Features} & \multicolumn{2}{c|}{\textbf{MVSA-Single}} & \multicolumn{2}{c|}{\textbf{MVSA-Multiple}} \\ \hline
                  & \textbf{ACC}                 & \textbf{F1}                  & \textbf{ACC}                  & \textbf{F1}                   \\ \hline
$E_T^{RAW}$         &       68.46         &      66.01         &       62.83          &    57.70                  \\ \hline
$E_T^{+HT}$         & 68.88               & 66.49              &        60.09         &     55.58                 \\ \hline
$E_T^{-HT}$         &      67.50	          &    64.50            &       65.65          &    \textbf{59.61}                \\ \hline
$C_T^{+HT}$              & \textbf{71.00}      & \textbf{68.47}      &   58.52    &   54.50   \\ \hline
$C_T^{-HT}$              &   68.72    &  65.59   &  \textbf{65.70}    &    59.43      \\ \hline
\hline
$E_o$               & 64.69               & 61.40               &        65.28         &      56.63        \\ \hline
$E_s$               & 64.66               & 61.51               &         65.13       &     56.00              \\ \hline
$E_a$             & 64.89               & 61.65               &        64.85        &      56.02               \\ \hline
$E_f$              &       59.63         &     48.48           &       \textbf{66.41}         &      53.21                 \\ \hline
$C_I$              &    \textbf{72.09}    &   \textbf{70.03}    &   65.42     &  \textbf{59.22}       \\ \hline
\hline
$C_I+E_o+E_s+E_f$      &         \textbf{70.29}            &      \textbf{69.51}                &        63.65             &     59.87                  \\ \hline
$C_I+E_a+C_f$ &     69.70 & 68.66   &   \textbf{63.79}   & \textbf{60.33} \\ \hline
$E_o+E_s+E_a+E_f$      &       66.42        &     65.40         &    63.49       &  58.58 \\ \hline
$E_o+E_a+E_f$      &      66.10       &      65.05        &     63.75      &  58.89 \\ \hline
\end{tabular}\label{tab:unimodal_results}
\end{table}
\egroup

\subsection{Results}
As listing all feature combinations is ineffective, we report the ones which are the most informative and reflect the use of each feature type. Also, we only show a maximum combination with four types of features (visual + textual) as no considerable improvement was observed with five features.
\subsubsection{Unimodal Results:}
Table~\ref{tab:unimodal_results} presents the evaluation results of unimodal textual and visual features for the two datasets. For \emph{MVSA-S}, we found out that including hashtag words in the text ($E_T^{+HT}/C_T^{+HT}$) gives slightly better performance than removing hashtags or using raw tweet ($E_T^{RAW}$) text. On the other hand, excluding hashtag words ($E_T^{-HT}/C_T^{-HT}$) from the text works better for the larger \emph{MVSA-M}, where the inclusion of hashtags degrades the average accuracy by almost \num{6}\%. The $E_T^{RAW}$ performance is slightly better than $E_T^{+HT}$ for \emph{MVSA-M} as \emph{RoBERTa's}~\cite{liu2019roberta} word piece tokenization tokenizes hashtags differently than the pre-processing we used in $E_T^{+HT}$. This also shows that for tasks like sentiment detection, pre-processing noisy tweet text can be very crucial. For single visual-only models, we can see that all the visual features except facial expressions ($E_f$) on their own are useful for sentiment detection. For both modalities, \emph{CLIP} features ($C_I$ and $C_T$) outperform all the other features by 2 to 6\% for \emph{MVSA-S} and show similar or slightly better results for \emph{MVSA-M}. This suggests that the pre-training strategy used in \emph{CLIP} learns expressive visual and textual features which can be used in multimodal downstream tasks. Interestingly, $C_I$ outperforms all other unimodal features for both datasets. \\
\subsubsection{Visual Combination Results:} Adding any other feature with $C_I$ degrades the performance indicating that other visual features are not compatible with $C_I$ in our model, although they slightly increase F1 for \emph{MVSA-M}. With other visual features, we see that the addition of each type of feature (like $E_o+E_a+E_f$) increases the performance in both datasets, especially increase in accuracy and F1 measure by \num{1}\% to \num{4}\% on \emph{MVSA-S}. The improvement can be attributed to emotion features $E_a$ and $E_f$, which shows that the affective image content is equally important in addition to object and scene information. Facial features on their own perform the worst (very low F1) across datasets as almost \num{50}\% of the images have no detectable faces. For \emph{MVSA-M} in particular, the combination of visual modalities only increases the F1 score and needs further analysis.

\bgroup
\setlength{\tabcolsep}{6pt} 
\renewcommand{\arraystretch}{1.2} 
\begin{table}[!ht]
\caption{Multimodal results for \emph{MVSA-Single} and \emph{MVSA-Multiple}. Accuracy and F1 scores are averaged over 10 folds.}
\begin{tabular}{|l|c|c|c|c|}
\hline
\textbf{Features} & \multicolumn{2}{c|}{\textbf{MVSA-Single}}     & \multicolumn{2}{c|}{\textbf{MVSA-Multiple}}   \\ \hline
                  & \textbf{ACC}                   & \textbf{F1}                    & \textbf{ACC}                   & \textbf{F1}                    \\ \hline
                  & \multicolumn{2}{l|}{$E_T^{+HT}$}                  & \multicolumn{2}{l|}{$E_T^{-HT}$}                  \\ \hline

$E_o$      &   71.80    &   70.09    &   66.28    &  60.98     \\ \hline
$E_s$      &   72.53    &   70.77    &   65.80    &  60.59     \\ \hline
$E_a$      &    71.98   &   70.20     &  66.27     &  61.13     \\ \hline
$E_o+E_f$         & 72.85 & 71.57 & 66.01 & 62.51 \\ \hline
$E_s+E_a$         &    72.80          &     71.32         &      66.12        &    61.49          \\ \hline
$E_o+E_a+E_f$         &     72.93         &   71.80           &     66.31        &    \textbf{62.76}          \\ \hline
$E_s+E_a+E_f$         &      72.93        &    71.69          &     66.19        &     62.57         \\ \hline
$C_I$      &  \textbf{75.33} & 73.76 &  \textbf{66.35} & 61.89  \\ \hline
$C_I+E_f$           & 75.00          &   \textbf{73.96} & 66.08  & 62.52  \\ \hline
$C_I+E_a$           &  73.95          &     72.86  & 65.18  & 62.03  \\ \hline
$C_I+E_s+E_f$    & 74.73  & 73.60 & 66.02 & 62.51 \\ \hline
\hline
& \multicolumn{2}{l|}{$C_T^{+HT}$}                  & \multicolumn{2}{l|}{$C_T^{-HT}$}                  \\ \hline
$C_I$     &       \textbf{74.97}       &       \textbf{73.32}       &        \textbf{66.09}           &  61.27         \\ \hline
$C_I + E_f$     &    74.00          &   72.58           &        65.32           & \textbf{61.34}         \\ \hline
$C_I + E_a$     &    73.29          &     72.15         &        64.68           &  61.13         \\ \hline
$C_I+E_s+E_f$    & 73.89  & 72.68 & 65.43 &  61.49 \\ \hline
\end{tabular}\label{tab:multimodal_results}
\end{table}
\egroup

\subsubsection{Multimodal Results}
Finally, the combination of two modalities increases the sentiment prediction performance across the splits and considerably increases (by \num{4}\%) both measures on \emph{MVSA-S} as shown ($C_I+E_T^{+HT}$) in Table~\ref{tab:multimodal_results}. This improvement can be seen across all the splits for average, minimum, and maximum values in Table~\ref{tab:comparison}, and shows that the addition of modalities not only increases the best score but works for most of the splits. For the \emph{MVSA-M}, the improvement is minimal with \num{1}\% accuracy and \num{3}\% F1 from unimodal models. Interestingly, visual features combined with \emph{RoBERTa} ($E_T^{+HT}/E_T^{-HT}$) pooled features always outperform combinations with \emph{CLIP's} text features ($C_T^{+HT}/C_T^{-HT}$) as some are shown in two separate grouped blocks in Table~\ref{tab:multimodal_results}. 
This limited performance can be attributed to three issues: 1) considerably higher number of neutral samples that have a higher chance of getting classified as negative or positive, 2) the label pooling strategy~\cite{xu2017multisentinet} used to pool labels from three annotators, which gives preference to positive or negative over the neutral label (refer above) and results in a larger number of disputable labels, and 3) the model's inability to capture the interactions and differentiate between neutral and polar samples. In the next section, we conduct an error analysis of misclassified samples and group these errors for further consideration. 
Also, the combination of all four visual modalities in combination with text features did not improve the performance, possibly due to the increase in network parameters.

\bgroup
\setlength{\tabcolsep}{2pt} 
\renewcommand{\arraystretch}{1.2} 
\begin{table}[!ht]
\centering
	\caption{Comparison Results for \emph{MVSA-Single} and \emph{MVSA-Multiple}. Results marked with $^{\dagger}$ are taken from Jiang \textit{et. al.}~\cite{jiang2020fusion}, and $^*$ are results of re-implemented models.}
    \begin{tabular}{|l|c|c|c|c|c|c|}
    \hline
    \multicolumn{1}{|c|}{\textbf{Baseline}} & \multicolumn{3}{c|}{\textbf{MVSA-Single}} & \multicolumn{3}{c|}{\textbf{MVSA-Multiple}} \\ \hline
    & \multicolumn{2}{c|}{\textbf{ACC}}          & \multicolumn{1}{c|}{\textbf{F1}} &   \multicolumn{2}{c|}{\textbf{ACC}}          & \multicolumn{1}{c|}{\textbf{F1}}         \\ \hline
		\textbf{\begin{tabular}[c]{@{}l@{}}SentiBank \&\\ SentiStrength\end{tabular}}$^{\dagger}$~\cite{borth2013large}  & \multicolumn{2}{c|}{52.05}           & 50.08                &  \multicolumn{2}{c|}{65.62}           & 55.36          \\ \hline
		\textbf{CNN-Multi}$^{\dagger}$~\cite{cai2015convolutional}                                                 & \multicolumn{2}{c|}{61.20}           & 58.37                &  \multicolumn{2}{c|}{66.30}           & 64.19          \\ \hline
		\textbf{DNN-LR}$^{\dagger}$~\cite{yu2016visual}                                                               & \multicolumn{2}{c|}{61.42}           & 61.03                & \multicolumn{2}{c|}{67.86}           & 66.33          \\ \hline
		\textbf{MultiSentiNet}$^{\dagger}$~\cite{xu2017multisentinet}                                                         & \multicolumn{2}{c|}{69.84}          & 69.63                &  \multicolumn{2}{c|}{68.86}           & 68.11          \\ \hline
		\textbf{CoMN(6)}$^{\dagger}$~\cite{xu2018co}                                                              & \multicolumn{2}{c|}{70.51}           & 70.01               &  \multicolumn{2}{c|}{70.57}           & 70.38          \\ \hline
		\textbf{FENet-BERT}$^{\dagger}$~\cite{jiang2020fusion}                                                           & \multicolumn{2}{c|}{74.21}           & 74.06                & \multicolumn{2}{c|}{71.46}          & 71.21          \\ \hline
		\hline
		
    \textbf{Models} & \multicolumn{3}{c|}{\textbf{ACC}}          & \multicolumn{3}{c|}{\textbf{F1}}           \\ \hline
    & \textbf{Avg} & \textbf{Min} & \textbf{Max} & \textbf{Avg} & \textbf{Min} & \textbf{Max} \\ \hline
     & \multicolumn{6}{c|}{\textbf{MVSA-Single}}           \\ \hline
    \textbf{MultiSentiNet}*           &      63.27        &     57.87         &      69.25        &     59.12         &       57.83       &   63.61           \\ \hline
         \textbf{FENet-BERT}*           &      69.02        &     63.76         &       71.67       &      67.30        &      61.42        &   69.97           \\ \hline
         \textbf{Se-MLNN($C_I+E_T^{+HT}$)}         &      \textbf{75.33}        &     \textbf{70.51}         &       \textbf{82.04}      &      \textbf{73.76}       &      \textbf{69.82}        &   \textbf{81.14}           \\ \hline
         & \multicolumn{6}{c|}{\textbf{MVSA-Multiple}}           \\ \hline
         \textbf{MultiSentiNet}*           &      63.08        &     54.32         &      67.10        &     59.12         &       54.43       &   58.57           \\ \hline
         \textbf{FENet-BERT}*           &      \textbf{68.61}        &     \textbf{61.47}         &       \textbf{74.40}      &      \textbf{65.80}       &      \textbf{60.84}        &   \textbf{73.56}           \\ \hline
         \textbf{Se-MLNN($C_I+E_T^{-HT}$})           &      66.35        &     59.54         &       70.27      &      61.89       &      55.43        &  65.33           \\ \hline
    \end{tabular}\label{tab:comparison}
\end{table}
\egroup

\subsection{Error Analysis and Discussion}
\begin{figure*}[!ht]
    
     \centering
     \includegraphics[width=1.0\textwidth]{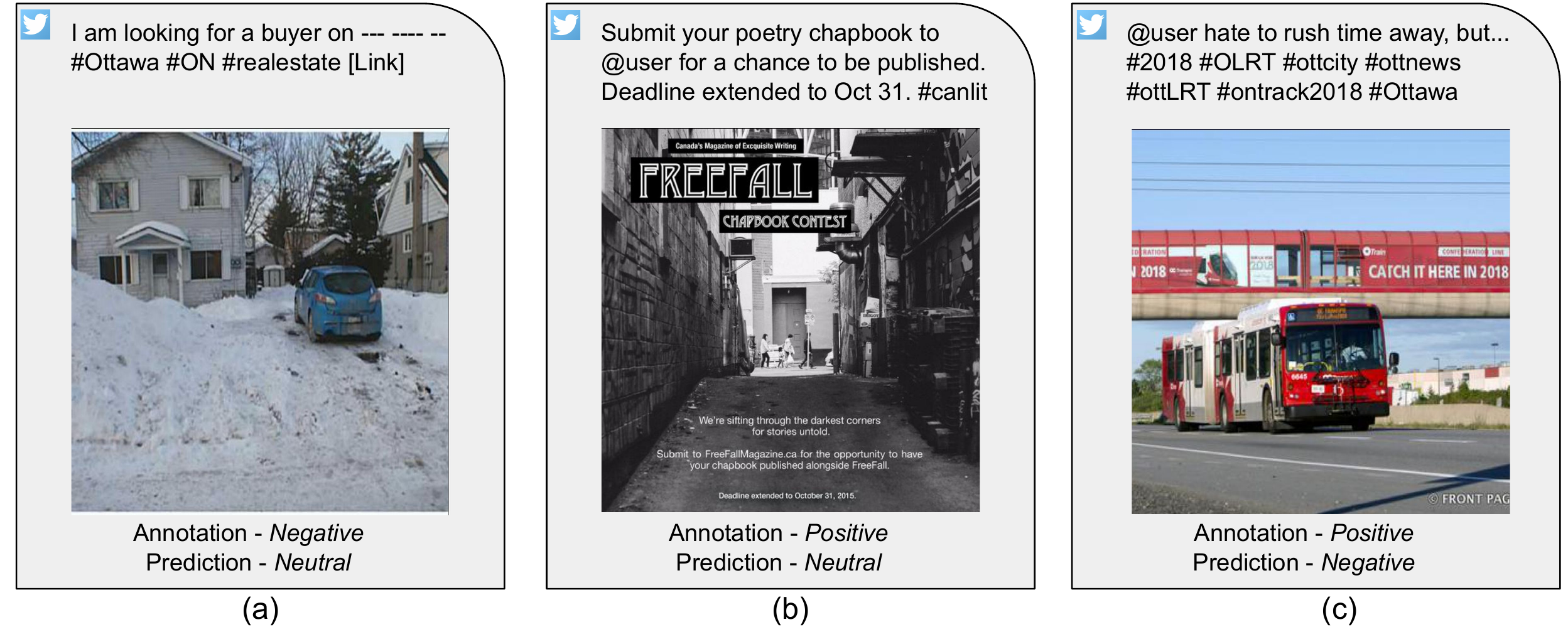}
     \caption{Samples from \emph{MVSA-Multiple} showing three errors}
     \label{fig:samples}
\end{figure*}
\subsubsection{Error Analysis}\label{sec:error}
As stated above, the presence or absence of hashtags has an adverse effect on the performance of both datasets. Further analysis revealed that hashtags that overlap between train, validation, and test splits in \emph{MVSA-S} reflect emotions, and the top-performing splits have a higher overlap of such hashtags. For instance, top overlapping hashtags across different splits in the dataset are words like \textit{love, happy, passionate, winter, positive, wild, strong, depressed, calm, excited, broken, fear, zealous, joy}, and \textit{fun}. On the other hand, despite higher overlap between hashtags in \emph{MVSA-M}, the hashtags consist of neutral words like \textit{toronto, elxn42, yyc, cdnpoli, nationaldogday, job, vancouver, canada, music, ottawa, hiring, realestate, realchange, halifax, photography} and \textit{food}. Although hashtags provide additional context and may reflect sentiment on social media posts, it is important to understand whether they should be an additional modality to text, filtered, or considered part of the text to better incorporate them in the model. On further inspection of both the datasets, we found around 500 and 700 non-English tweets in \emph{MVSA-S} and \emph{MVSA-M} respectively. Such tweets should either be removed from the datasets or multi-lingual models should be considered for language agnostic sentiment detection.
In order to understand the errors from both image and text perspective, we analyze 150 tweets (50 from each class) from the \emph{MVSA-M} that were incorrectly classified by our best model. We divide the errors into three categories. 

\textbf{Disputable:} We found a considerable amount of disputable labels across all three classes, where some labels are highly disputable and open for discussion. We consider 70 out of 150 labels to be open for discussion, where the majority of them are from positive and negative classes. Quite a few tweets that are disputable in the negative class are selfies of individuals smiling with no negative word mentions in the text. 
In the positive class, a recurring trait in tweets includes advertisements with no positive connotation in either text or image and thus should be labeled as neutral. 
For example, both tweet text and image in Figure~\ref{fig:samples}a have neither positive nor negative connotations but it is annotated as \textit{negative}. \\
\textbf{Text in Image:} We also found 66 errors that were due to the embedded text in the image, for which we have not incorporated a visual model in the sentiment prediction. Other common types of images in this category are memes, chat snapshots, weather reports, advertisements, and graphs.
For example, text content in the image (see Figure~\ref{fig:samples}b) has important details of a contest that could reflect or change the sentiment of the tweet. \\
\textbf{Missing Context:} Around 67 errors fall into this category where background or cultural information is required to connect the content from both text and image. 
For example, the tweet text in Figure~\ref{fig:samples}c has a negative connotation, but the image and hashtags refer to city's bus service among other things that require additional context. Errors, where there is no contextual overlap of image and text or the information is very abstract, are also grouped here. 

\subsubsection{Quantitative Examples}

\begin{figure*}[!h]
    
     \centering
     \includegraphics[width=1.0\textwidth]{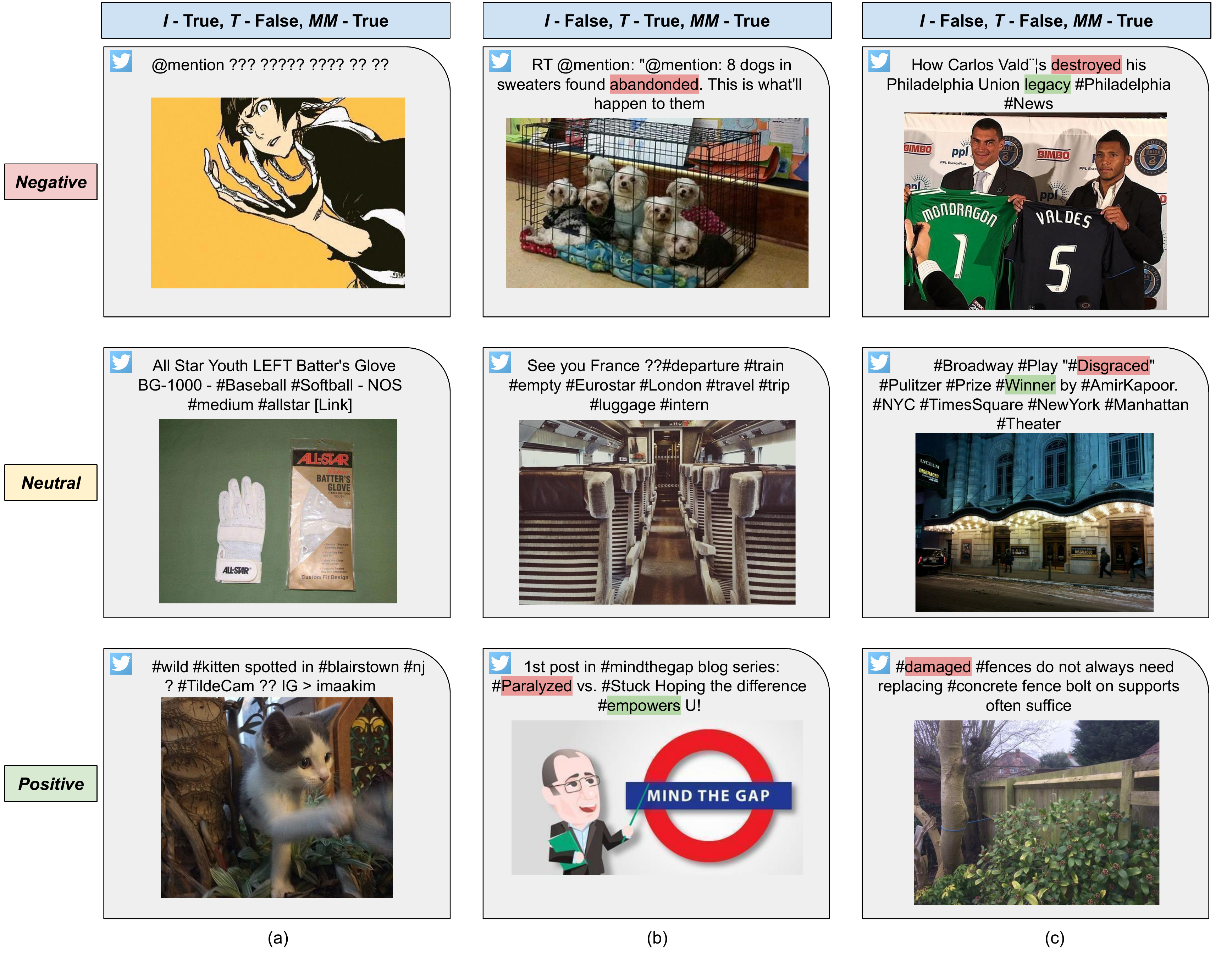}
     \caption{Examples from \emph{MVSA-Single} where multimodal (\textit{MM}) models predict the correct label and either image-only (\textit{I}) or text-only (\textit{T}) models fail}
     \label{fig:quantitative}
\end{figure*}

In Figure~\ref{fig:quantitative}, we show a few examples from \emph{MVSA-S} where the multimodal model predicts the correct label and unimodal models make an incorrect classification. In the first column (a), where image-only and text-only models predict the correct and incorrect class respectively, texts are mostly neutral and images contain the sentiment specific information. For instance graphic skeletal hand and a cat in the negative and positive example respectively. Some other examples in this category with polar sentiment include graphic images of dead animals and selfies of people with prominent facial expressions. In the neutral category, there are several images of objects (see middle row in Figure~\ref{fig:quantitative}~a) and advertisements with neutral text. In the second column (b), where image-only and text-only models predict the incorrect and correct class respectively, texts have negative or positive connotation and both modalities need to be seen to predict the correct label. For instance, the negative pair in column (b) has an image of puppies in a cage which does not have a negative connotation, but the text gives important context that the puppies are abandoned. Similar case can be made for the positive pair in the same column. In the last column (c), where both unimodal models predicted the incorrect class, in addition to a few valid cases, there are examples of disputable labels as noted in the previous section~\ref{sec:error}. For instance, the positive pair in column (c) has no positive connotation in either text or image and should have a ground truth label as neutral.

\section{Conclusions and Future Work}\label{sec:conclusion}
In this paper, we have 
presented a comprehensive experimental evaluation on visual, textual, and multimodal features for sentiment prediction of (multimodal) tweets, including the recently introduced CLIP embeddings. Furthermore, we have compared the performance with six state-of-the-art methods. It turned out that CLIP embeddings can serve as a powerful baseline for the task of multimodal sentiment prediction in tweets.
Unlike the used evaluation methodology in previous work, we have introduced a fair and reproducible experimental setup with a 10-fold cross-validation 
that hopefully provides a useful benchmark for future research and comparison. 
In future work, we will take cues from the error analysis and focus on models that encode textual content in images as well as contextual information in visual and textual modalities.

\section*{Acknowledgements}
This work was funded by European Union’s Horizon 2020 research and innovation programme under the Marie Skłodowska-Curie grant agreement no 812997.
\bibliographystyle{ACM-Reference-Format}
\bibliography{main}

\end{document}